# Correlation between thermal expansion and heat capacity

Jozsef Garai

*Department of Earth Sciences, Florida International University*
Tel: 1-786-247-5414
fax: 1-305-348-3877
*E-mail address*: jozsef.garai@fiu.edu

**Abstract**

Theoretically predicted linear correlation between the volume coefficient of thermal expansion and the thermal heat capacity was investigated for highly symmetrical atomic arrangements. Normalizing the data of these thermodynamic parameters to the Debye temperature gives practically identical curves from zero Kelvin to the Debye temperature. This result is consistent with the predicted linear correlation. At temperatures higher than the Debye temperature the normalized values of the thermal expansion are always higher than the normalized value of the heat capacity. The detected correlation has significant computational advantage since it allows calculating the volume coefficient of thermal expansion from one experimental data by using the Debye function.

*Keywords:* Thermal Properties; Thermal expansion; Heat capacity; Correlation; Highly symmetrical atomic arrangements

## 1. Introduction

The volume coefficient of thermal expansion $[\alpha_V]$ is described as

$$\alpha_V = \frac{1}{V}\left(\frac{dV}{dT}\right)_p \tag{1}$$

where V is the volume, T is the temperature, and P is the pressure. The volume coefficient of thermal expansion is temperature dependent requiring numerous experiments for its complete description. The experiments are time consuming and technically difficult at extreme temperatures.

The theory of thermal expansion is well defined. The volume expansion can be determined by calculating the anharmonic term of the lattice vibration [ex. 1, 2]. This traditional approach has been challenged by a simple classical model [3, 4] which gave good and excellent agreement with the experimental data of Ne, Ar, Kr, and Xe. In these calculations it was assumed that the potential is Lennard-Jones and that the mean value of the interatomic atomic distance [$\langle R \rangle$] can be used to calculate the thermal expansion.

$$[\alpha(T)]_{P=0} = \frac{1}{R_0}\left[\frac{d\langle R(T)\rangle}{dT}\right]_{P=0} \qquad (2)$$

The mean interatomic distance with a good approximation is proportional to the mean vibrational energy per atom in a solid [2]. The thermal energy of a system [$Q^{thermal}$] is the product of the mean vibrational energy of an atom and the number of atoms in the system. The zero-term energy of the atoms is not included in the thermal energy. Introducing a constant [$a$] allows to substitute the mean interatomic distance with the thermal energy of a system. Equation (2) can be rewritten then

$$[\alpha(T)]_{P=0} = \frac{a}{n}\left[\frac{dQ^{thermal}(\langle R(T)\rangle)}{dT}\right]_{P=0} \qquad (3)$$

where n is the number of moles.

The molar thermal heat capacity $c(\varphi)^{thermal}$ is defined as:

$$c(\varphi)^{thermal} = \frac{1}{n\delta T}\int_T^{T+\delta T}\delta Q(\varphi)^{thermal} = \frac{1}{n}\frac{dQ(\varphi)^{thermal}}{dT} \qquad (4)$$

where $\varphi = g(gas), s(solid), l(liquid)$. Combining equation (3) and (4) gives

$$[\alpha_V(T)]_{P=0} = a\left[c(s)^{thermal}\right]_{P=0}. \qquad (5)$$

The molar thermal heat capacity is pressure independent. The pressure has effect on the equilibrium separation [$R_0$] of the atoms, which is incorporated in the constant. The pressure effect on $R_0$ can be accommodated by assuming that the introduced constant [a] is pressure dependent.

$$[\alpha_V(T)]_P = a_P \left[c(s)^{thermal}\right] \tag{6}$$

The pressure up to few GPa has minor effect on the heat capacity [5]. In this pressure range the molar thermal heat capacity can be replaced with the constant pressure molar heat capacity.

$$[\alpha_V(T)]_P = a_P [c(s)]_P \tag{7}$$

Equation (7) predicts linear correlation between the volume coefficient of thermal expansion and the molar heat capacity at constant pressure. This correlation will be considered in detail.

## 2. Correlation between the volume coefficient of thermal expansion and molar heat capacity

Experimental data of highly symmetrical mono-atomic arrangements [6] Ag, Al, Au, Ba, Co, Cr, Cu, Fe, K, Ni, Pb, Pt, Ti, V have been used to investigate the predicted correlation between the volume coefficient of thermal expansion and the molar heat capacity at atmospheric pressure. The experimental values of these two thermodynamic parameters were not necessarily determined at the same temperature. In order to make these parameters comparable, the two data sets were normalized to the Debye temperature. It has been found that the normalized values of the volume coefficient of thermal expansion and the heat capacities are practically identical below the Debye temperature. Above the Debye temperature the normalized values of the volume coefficient of thermal expansion are higher than the normalized values of the heat capacity. The characteristic behavior is shown on Fig. 1. These results are interpreted as follows. The identical curves of the normalized values of the volume coefficient of thermal expansion and molar heat capacity at atmospheric pressure confirm the predicted linear correlation between these thermodynamic parameters from zero Kelvin to the Debye temperature. The higher normalized values of the volume coefficient of thermal expansion above the Debye temperature might be the result of the additional expansion caused by lattice vacancies [7, 8].

## 3. Computational advantages

Employing the detected linear correlation between the volume coefficient of thermal expansion and the molar thermal heat capacity allows calculating the volume coefficient of thermal expansion from any experiment. Using equation (6) for two different temperatures and dividing equation one by equation two gives

$$\left[\frac{\alpha_V(T_1)}{\alpha_V(T_2)}\right]_P = \frac{c(s)_{T_1}^{thermal}}{c(s)_{T_2}^{thermal}} \quad \text{and} \quad [\alpha_V(T_2)]_P = \frac{[\alpha_V(T_1)]_P}{c(s)_{T_1}^{thermal}} c(s)_{T_2}^{thermal}. \tag{8}$$

The pressure effect is canceled out in equation (8) by the division. The molar thermal heat capacity can be calculated by using the Debye function [9].

$$c(s)^{thermal} \approx c(s)^{Debye} = 3Rf \qquad f = 3\left(\frac{T}{T_D}\right)^3 \int_0^{x_D} \frac{x^4 e^x}{(e^x - 1)^2} dx \tag{9}$$

and

$$x = \frac{h\omega}{2\pi k_B T} \quad \text{and} \quad x_D = \frac{h\omega_D}{2\pi k_B T} = \frac{T}{T_D} \tag{10}$$

where h is the Planck's constant, $\omega$ is the frequency, $\omega_D$ is the Debye frequency and $T_D$ is the Debye temperature. This equation has to be evaluated numerically [10].

Using the experimental value of the volume coefficient of thermal expansion at the Debye temperature $[\alpha(T_D)]$ the volume coefficient of thermal expansion was calculated between zero Kelvin and the Debye temperature. The calculated and the experimental values are plotted in Fig. 2. Based on visual inspection the correlations are excellent for elements (Ag; Al; Au; Cu;

Ni; Pb; Pt) with face centered cubic structure, very good for Co and Ti with hexagonal close packed structure and good for element (Ba; Cr; Fe; K; V) with body centered cubic structure.

## 4. Conclusions

The theoretically predicted linear correlation between the volume coefficient of thermal expansion and the heat capacity has been confirmed for highly symmetrical mono-atomic arrangements.

The detected correlation allows calculating the volume coefficient of thermal expansion from an experiment conducted at a preferably chosen temperature.

## References

[1] C. Kittel, Introduction to Solid State Physics 6$^{th}$ edition, New York, Wiley, 1986, p. 114.
[2] R.A. Levy, Principles of Solid State Physics, New York, Academic, 1968, p. 141.
[3] P. Mohazzabi and F. Behroozi, Phys. Rev. B 36 (1987) 9820-9823.
[4] P. Mohazzabi and F. Behroozi, Eur. J. Phys. 18 (1997) 237-240.
[5] J. Garai (2005) http://arxiv.org/physics/0511066.
[6] Handbook of Physical Quantities, edited by I. S. Grigoriev and E. Z. Meilikhov, CRC Press, Inc. Boca Raton, FL, USA, 1997.
[7] R.O. Simmons and R.W. Balluffi, Phys. Rev. 117 (1960) 52-61.
[8] Th. Hehenkamp, W. Berger, J.–E. Kluin, Ch. Lüdecke, and J. Wolff, Phys. Rev. B 45 (1992) 1998-2003.
[9] P. Debye, Ann. Physik 39, (1912) 789.
[10] Landolt-Bornstein, Zahlenwerte und Funktionen aus Physic, Chemie, Astronomie, Geophysic, und Technik, II. Band, Eigenschaften der Materie in Ihren Aggregatzustanden, 4 Teil, Kalorische Zustandsgrossen, Springer-Verlag, 1961.

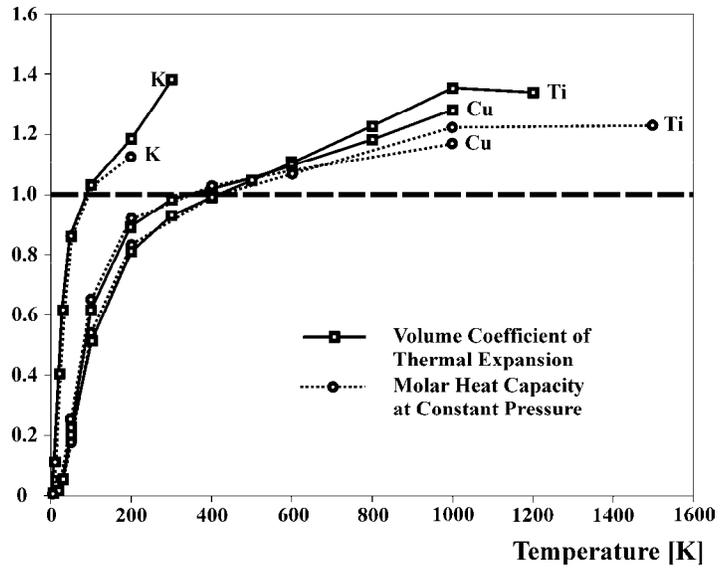

Fig. 1. Normalized values of the volume coefficient of thermal expansion and the molar volume heat capacity at constant pressure as function of temperature for K, Cu, and Ti.

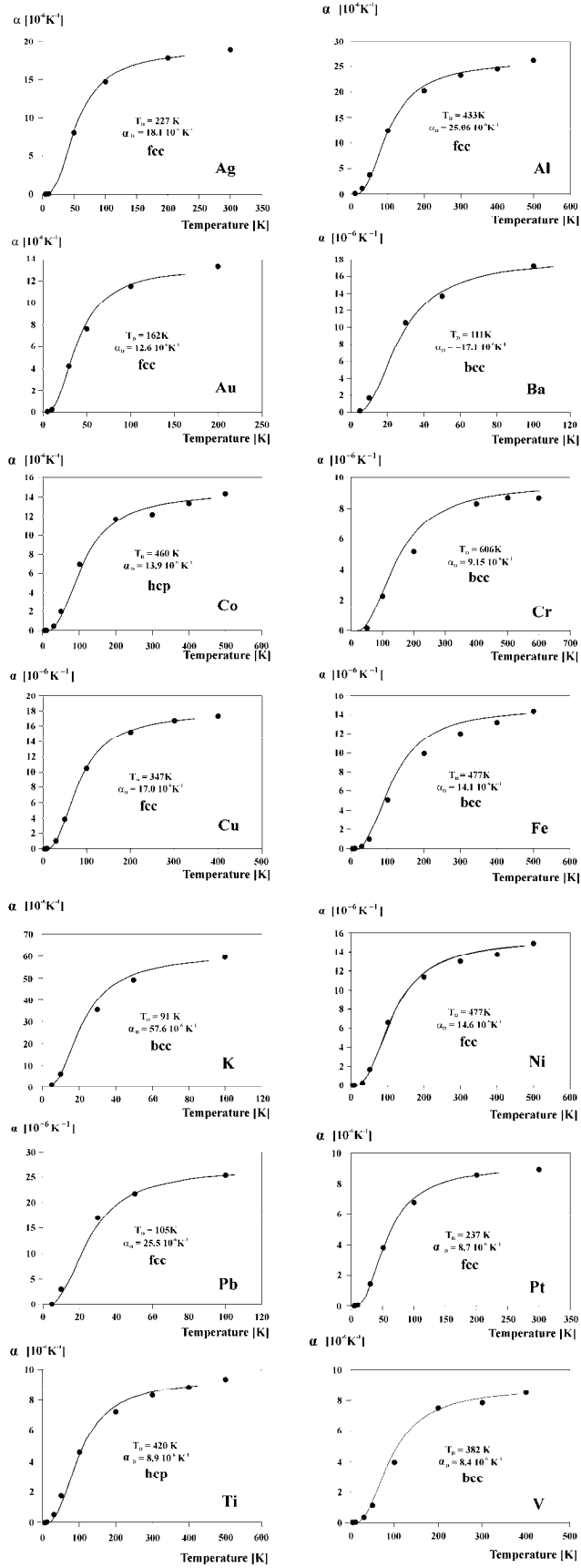

Fig. 2. The solid lines are the calculated volume coefficients of thermal expansion while the dots represent the experimental values [6].